\documentclass[prd,preprintnumbers,nofootinbib,aps,9pt]{revtex4}
\usepackage{subeqnarray}
\usepackage{epsfig,amsmath,amssymb}
\usepackage{mathrsfs}
\usepackage[usenames,dvipsnames]{color}
\usepackage[pagebackref=true, colorlinks=true]{hyperref}
\definecolor{redish}{rgb}{0.7,0.2,0.0}  
\definecolor{bluish}{rgb}{0.2,0.5,0.8}
\hypersetup{linkcolor=redish,          
                  citecolor=blue,        
                  filecolor=magenta,      
                  urlcolor=bluish}          

\DeclareFontFamily{U}{rsfs}{}         
\DeclareFontShape{U}{rsfs}{m}{n}{<5> rsfs5 <6><7> rsfs7          %
  <8><9><10><10.95><12><14.4><17.28><20.74><24.88> rsfs10}{}     %
\DeclareMathAlphabet{\mathfs}{U}{rsfs}{m}{n}                     %

\newcommand{\ba}{\nopagebreak[3]\begin{eqnarray}}
\newcommand{\ea}{\end{eqnarray}}
\newcommand{\bii}{\begin{itemize}}
\newcommand{\eii}{\end{itemize}}

\def \({\left(}
\def \){\right)}
\def \[{\left[}
\def \]{\right]}
\begin{document}
\title{Singularity from star collapse, torsion and asymptotic safety of gravity}
\author{Abhishek Majhi}%
\email{abhishek.majhi@gmail.com}
\affiliation{
Instituto de Ciencias Nucleares\\
Universidad Nacional Autonoma de Mexico\\
A. Postal 70-543, Mexico D.F. 04510, Mexico\\
}

\begin{abstract}
A star of mass greater than a critical mass is believed to undergo  gravitational collapse to form a singularity, owing to Hawking-Penrose singularity theorem which is based on the Raychaudhuri equation in the absence of torsion. We argue that the spin-aspect of matter can lead to the evasion of singularity,  caused by its mass-aspect,  via torsion in asymptotically safe gravity. 
\end{abstract}
\maketitle

A star of mass $(M)$ greater than a critical mass $(M_c)$ can not withstand  its own gravity by means of the degeneracy pressure due to the Pauli exclusion principle obeyed by the constituent fermion-matter of the star (e.g. see \cite{shapiro}). As a result, it collapses under its own gravity and forms a black hole of mass $M$, which is characterized by the presence of a horizon. Then, it is believed that this collapsing matter, which we shall assume to consist of fermions, continues all the way to end up in a singularity. This belief is mathematically supported by the Hawking-Penrose singularity theorem which leads to incompleteness of time-like and null geodesics via Raychaudhuri equation in Einstein's general relativity (EGR) \cite{hp69}. However, if we carefully look at the aspects of theoretical physics that may possibly be involved in this star-collapse situation, along with the most conservative approach to quantum gravity, namely The Asymptotic Safety program \cite{reuter}, 
we shall find  enough reason to doubt our conclusion about the singularity formation.

  The methodology used to estimate $M_c$ is based on the application of quantum mechanics of a collection (gas/fluid) of free fermions; the energy-momentum tensor of such a fermionic gas is considered along with EGR leading to an equation of state \cite{ns1,ns2}. However, the fact that the constituent particles are fermions is a very subtle issue.  It is a well known fact that the proof of the spin-statistics theorem is tied to Lorentz symmetry in flat spacetime \cite{sud}. While a proper mathematical proof for spin-statistics theorem on a generic curved spacetime is yet to be penned \cite{few}, it can not be denied that the constituents of a collapsing star live on a curved spacetime. Added to this, the notion of `particles' become illusive on a curved spacetime \cite{parker}. Therefore what is done while calculating $M_c$ is not  accurate from the theoretical point of view. Instead of quantum field theory on a curved spacetime (QFTCS),  the methodology is based on our intuitive approximation of the possible situation resulting in a hybrid mixture of apparently disconnected notions of non-relativistic quantum mechanics and EGR \cite{ns1,ns2}. In fact, the use of QFTCS is also inadequate to handle the situation at hand.
This is because, in the process of applying QFTCS to the star-collapse scenario, we are neglecting the back reaction of the matter on geometry. This is tantamount to ignore a major part of the physics in this particular situation. QFTCS is only well-suited when the back-reaction on the geometry can be neglected and the matter can be treated as test fields propagating on a fixed background. 
  
  Therefore, it seems that a more appropriate method is to apply the full dynamics of both the geometry and the matter sectors together so as to have a precise theoretical description of the star-collapse situation. At the classical level, this implies that we have to write down an action of gravity coupled to fermions which occurs naturally in the Einstein-Cartan-Sciama-Kibble (ECSK) theory rather than in EGR \cite{hehl}. Hence, it seems that ECSK theory provides a natural stage for probing the dynamics of the constituent matter of a collapsing star. Once we accept this from the theoretical point of view, then there is an interesting consequence that follows inevitably. In EGR we do not have torsion. However, in ECSK theory we have torsion and it is sourced by the spin-density of the fermions \cite{hehl}. Therefore,  presence of torsion seems to be inevitable in the collapsing matter of a star according to this theory. 
  
  Now, the singularity theorem was proved using Raychaudhuri equation in the absence of torsion \cite{hp69}. The presence of torsion radically changes the scenario \cite{hehl}. Two classes of preferred curves come into the picture, namely the autoparallel curves which are the straightest lines of the manifold and the extremal curves which are the shortest or longest lines of the manifold. This distinction vanishes in the absence of torsion and the two classes of curves merge to give the unique notion of geodesics. Interestingly, fermions free fall along the autoparallel trajectories and not the extremal curves. However, the singularity theorem was based on the extremal nature of geodesics \cite{hp69}.  Given this fundamental change in the scenario, although the Raychaudhuri equation has recently been investigated in the presence of torsion \cite{rc17}, the status of the singularity theorem  is still unknown. So, in the absence of a proper proof (or disproof) of the singularity theorem in the presence of torsion we are not in a position to conclude whether a star of $M>M_c$ will collapse to form a singularity or not. 

However, there is a catch in this line of arguments. 
Unfortunately, even though we see fermions in our everyday life, no signature of torsion has been detected till date because of  the very weak nature of the coupling of torsion to spin-density due to the very small value of the gravitational constant $(G)$\footnote{ Although the non-propagating nature of torsion is another possible reason for its undetectability, but it is not relevant in the present discussion.} \cite{hehl}.  Hence, on this ground alone the above arguments can be rejected.
So, the obvious question is now the following. If torsion is undetectable in our observable universe, how is it possible for torsion to have an effect on the collapsing matter of the star? The possibility comes from a very key element of theoretical physics, or more specifically quantum field theory, that we have not yet considered, namely, the running of coupling constants due to renormalization. The value of $G$ is generally measured in the low energy and weak gravity regime \cite{Gm}.  The renormalization group flow resulting in scale dependent coupling constants of the theory may drastically modify the effective value of $G$ in the situations that prevail inside a star which is on the verge of gravitational collapse and involves strong gravity and high energy physics. This viewpoint is due to the Asymptotic Safety approach to quantum gravity which pushes the known ideas of quantum field theory of matter fields to that of gravity \cite{reuter}.  The effect of matter on the running of $G$ is a subject of active research (see \cite{reuter} and the references therein). So, the possibility can not be ruled out that the value of $G$ may increase with the energy scale in the presence of fermions and consequently enhancing the effect of torsion in the collapsing matter. In fact, we have a theoretical hint of such a possibility of increment of $G$ in the presence of fermions \cite{grun}. Hence, there is a likelihood for torsion to play a significant role in the scenario under discussion.

Therefore, taking into account all the above discussed facts, it seems that there are seeds of reasonable doubt about the formation of singularity from gravitational collapse of a star with $M>M_c$ even if we remain within the domain of the most conservative approaches of theoretical physics. To mention, here we have assumed that although gravity overcomes the degeneracy pressure of the fermions for a star with $M>M_c$, it does not necessarily mean that the fermions lose their identity e.g. forming a hitherto unknown form of matter, etc. This assumption is required because the retention of their own identity by the fermions is a necessity for the presence of torsion which is sourced by the spin-density of the fermions. Finally, it is very tempting to conclude with the following comments. Mass and spin are two intrinsic properties of matter. They are obtained as the two Casimir invariants of the Poincar\'e symmetry group which is the  local gauge group associated with the ECSK theory \cite{hehl}. However, in EGR only the mass-aspect of matter is realized through the energy-momentum tensor and the theory predicts singularity.  Therefore, it seems that the completeness of the ECSK theory as a gauge theory may evade the singularity from a star-collapse by bringing in the spin-aspect of matter via torsion in asymptotically safe gravity.
\vspace{0.3cm}
\\ {\bf Acknowledgments:} The early stages of this work was supported by DGAPA postdoctoral fellowship of UNAM, Mexico. The author is grateful to Chandrachur Chakraborty and Parthasarathi Majumdar for useful suggestions and Cristobal Corral and Daryel Manreza Paret for helpful discussions.

\end{document}